\documentclass[prd,nofootinbib,aps,superscriptaddress,tightenlines,preprintnumbers,twocolumn]{revtex4}

\usepackage{epsfig,latexsym,amsmath,amssymb}

		\def\b{\beta}		
\def\d{\delta}		\def\e{\epsilon}		
		\def\h{\eta}		
\def\j{\psi}			\def\k{\kappa}		\def\l{\lambda}
\def\m{\mu}		\def\n{\nu}			\def\o{\omega}
\def\p{\pi}					
					
					\def\D{\Delta}
			\def\J{\Psi}
\def\L{\Lambda}			
\def\Q{\Theta}				
				
\def\bJ{\bar{\Psi}}
\def\bQ{\bar{\Theta}}

			\def\co{{\cal O}}

\def\lag{{\mathcal{L}}}

\def\half{{1 \over 2}}
\def\ki{{\bar{k}_i}}
\def\kj{{\bar{k}_j}}
\def\k3{{\bar{k}_3}}
\def\kpb{{\bar{k}_{\bot}}}
\def\kp{k_{\bot}}
\def\kpv{\vec{k}_{\bot}}
\def\pa{\partial}
\begin{document}

\preprint{\hbox{CALT-68-2669}  } 

\title{Classical stability of a homogeneous, anisotropic inflating space-time} 

\author{Timothy R. Dulaney}
\email[]{dulaney@theory.caltech.edu}
\affiliation{California Institute of Technology, Pasadena, CA 91125}

\author{Moira I. Gresham}
\email[]{moira@theory.caltech.edu}
\affiliation{California Institute of Technology, Pasadena, CA 91125}

\author{Mark B. Wise}
\email[]{wise@theory.caltech.edu}
\affiliation{California Institute of Technology, Pasadena, CA 91125}

\begin{abstract}
We study the classical stability of an anisotropic space-time seeded by a spacelike, fixed norm, dynamical vector field in a vacuum-energy-dominated inflationary era. It serves as a model for breaking isotropy during the inflationary era. We find that, for a range of parameters, the linear differential equations for  small perturbations about the background do not have a growing mode.  We also examine the energy of fluctuations about this background in flat-space. If the kinetic terms for the vector field do not take the form of a field strength tensor squared then there is a negative energy mode and the background is unstable. For the case where the kinetic term is of the form of a field strength tensor squared we show that perturbations about the background have positive energy at lowest order.
\end{abstract}

\date\today
\maketitle

\section{Introduction}

Inflation has become the standard paradigm for very early Universe cosmology. It supposes an era when the Universe was dominated by vacuum energy. During this era any classical  inhomogeneities are smoothed out by the exponential expansion  of the Universe. The inhomogeneities that we observe today originate in small quantum fluctuations that have their wavelength  redshifted  outside the horizon by the exponential expansion of the Universe.  At some time the Universe  exits the inflationary era and transitions to a radiation dominated era and then eventually to a matter dominated era.  After inflation, the Universe expands slower than the speed at which light travels.  This allows the perturbations with wavelengths that are outside of the horizon to eventually reenter the horizon and generate the inhomogeneities we observe in the microwave background (as well as the large scale structure of the Universe).

Since we have no direct probes of the inflationary era, we are compelled to consider the possibility that some tenets of physics---tenets that are fundamental to our current understanding of the Universe---only reflect post-inflationary developments. One such tenet is that  rotational invariance is not spontaneously (or explicitly) broken.  This symmetry implies angular momentum conservation and is necessary for the classification of elementary particles by their spin ({\it i.e}., their angular momentum in their rest frame).

Recently the possibility that rotational invariance is broken during the inflationary era has been studied \cite{Ackerman:2007, Gumrukcuoglu:2007bx, Gumrukcuoglu:2006xj, Pullen:2007tu, Ando:2007hc, Boehmer:2007ut}. In Ref.~\cite{Ackerman:2007} it was noted that if the breaking of rotational invariance is small there is a very simple and unique signature on the spectrum of density perturbations and hence on the anisotropy of the microwave background radiation.  Ackerman  {\it et.~al.}~also introduced a simple model for breaking rotational invariance during the inflationary era.  They calculated the influence of the breaking of rotational invariance on the spectrum of density perturbations, verifying the expectations based on their general arguments.   The Lagrange density for the model has the usual Einstein-Hilbert action, a cosmological constant term with vacuum energy density $\rho_{\Lambda}$ and the following  terms containing the  four-vector field $u^{\sigma}$,
\begin{eqnarray}
\label{ulag}
  {\cal{L}}_u &=&-\beta_1 \nabla^\mu u^\sigma \nabla_\mu u_\sigma - \beta_2 (\nabla_\mu u^\mu)^2 \nonumber \\  &-& \beta_3 \nabla^\mu u^\sigma \nabla_\sigma u_\mu + \lambda(u^{\mu} u_{\mu} - m^2)\ .
\end{eqnarray}
Here $\lambda$ is a Lagrange multiplier that enforces the constraint $ g_{\mu \nu} u^{\mu} u^{\nu}=m^2$. We take $m^2>0$ so the four-vector $u^{\sigma}$ is spacelike.

This model has a homogeneous but anisotropic background solution to the classical equations of motion. We choose the $x_3$-axis to be aligned along the  four-vector field,
\begin{equation}
u^0=0,~~u^1 = u^2=0~~\text{and}~~u^3 ={m \over b(t)},
\label{uvector}
\end{equation}
resulting in a space-time metric of the form,
\begin{equation}
\label{coord}
{\rm d}s^2=-{\rm d}t^2+a(t)^2{\rm d}{\bf x}_{\perp}^2+b(t)^2{\rm d}x_3^2.
\end{equation}
The breaking of rotational invariance by the four-vector field causes the $x_3$-direction to expand at a different rate than the $x_1$ and $x_2$ directions.  Explicitly,\footnote{It is assumed that the dynamics immediately preceding the inflationary era was rotationally invariant so that $a(0)=b(0)$.}
\begin{equation}
a(t)=e^{H_a t},~~~b(t)=e^{H_b t},
\label{spit1}
\end{equation}
where the Hubble parameters are related to the vacuum energy, the scale of rotational invariance breaking, and parameters in the vector Lagrangian by the following equations,
\begin{eqnarray}
\label{backgrounda}
&&H_a= {{\dot a}\over a} = H_b(1+16 \pi G  \beta_1 m^2),\nonumber  \\
&& H_b= {{\dot b}\over b} 
={{\sqrt{ 8 \pi G \rho_{\Lambda} \over (1+8 \pi G  \beta_1 m^2)(3+32
\pi G  \beta_1 m^2)}}} \,.
\end{eqnarray}
Notice that the background solution does not depend on the parameters $\beta_{2,3}$ and as $m \rightarrow 0$ the rate of expansion is the same for all spatial dimensions.

The purpose of this paper is to study the  classical stability of this solution.  We first consider the linear differential equations that result from expanding the classical equations of motion to linear order about the background solution discussed above. We find that, for small $G \beta_i m^2$,  small perturbations do not grow provided that $\beta_1 >0$ and $\beta_1+\beta_2 +\beta_3 \ge 0$.   Next we consider the energy of these solutions in flat-space.  Negative energy modes indicate an instability that might not show up in the analysis of the linearized equations of motion.

A special role is played by the case $\beta_1+\beta_2 +\beta_3= 0$ where, in flat-space, the kinetic terms for the vector field take the form of a field strength tensor squared.  In this case, we find that when the energy density is evaluated to quadratic order in  fluctuations about the background, all propagating modes have positive energy. However, when that equality is not satisfied, there is a mode that propagates with negative energy.  

Models with vector fields that spontaneously break Lorentz symmetry were first introduced in \cite{Kostelecky:1989jw}.  Similar fixed-norm timelike Lorentz-violating vector field models have been extensively studied; constraints required by theoretical and observational consistency have been placed on parameters in this theory \cite{Jacobson:2004ts, Carroll:2004ai, Lim:2004js, Eling:2003rd, Foster:2005dk, Li:2007vz, Seifert:2007fr, Foster:2006az, Eling:2006df, Jacobson:2000xp, Clayton:2001vy, Kostelecky:1989jw, Bluhm:2007bd}. For a review, see \cite{Jacobson:2008aj}. 

Others have classified gauge invariant perturbations in anisotropic scenarios  \cite{Pereira:2007yy, PhysRevD.34.3570}; however, we choose to work in a particular (nonstandard) gauge for calculational convenience.
 

\section{Strategy}


The effect of an isotropy-breaking vector field on the expansion of the Universe during inflation was recently studied \cite{Ackerman:2007}.  In this model, the four-vector $u^\mu$ is non-zero only during the
time interval $0<t<t_*$, where $t = 0$ is the beginning of inflation and $t_*$ is the end of inflation. We assume that the dynamics is rotationally invariant during reheating and
thereafter. During the time interval $0<t<t_*$, the dynamics of
interest in this paper  is governed by the action 
\begin{equation}
\label{dyn}
S=\int {\rm d}^4x {\sqrt{-g}}\left({1 \over 16 \pi G} R-\rho_{\Lambda} +  {\cal{L}}_u\right),
\end{equation}
where ${\cal L}_u $ is given in Eq.~(\ref{ulag}).
The homogeneous background  inflationary space-time solution is given in Eqs.~\eqref{coord}, \eqref{spit1} and \eqref{backgrounda}.

The energy-momentum tensor for $u^\mu$ derived from (\ref{ulag}) is \cite{Carroll:2004ai},
\begin{eqnarray}
\label{tmunu}
T_{\mu\nu}^{(u)}&=&2\beta_1(\nabla_{\mu} u^{\rho}\nabla_{\nu} u_{\rho}-\nabla^{\rho} u_{\mu} \nabla_{\rho} u_{\nu})  \nonumber \\
&&-2[\nabla_{\rho}(u_{(\mu}J^{\rho}{}_{\nu)})+\nabla_{\rho}(u^{\rho} J_{(\mu\nu)})-\nabla_{\rho}(u_{(\mu}J_{\nu)}{}^{\rho})] \nonumber \\
&&+2m^{-2}u_{\sigma}\nabla_{\rho} J^{\rho\sigma}u_{\mu} u_{\nu}+g_{\mu\nu}{\cal{L}}_u ,
\end{eqnarray}
where $J^{\mu}{}_{\sigma}$ is the current tensor,
\begin{equation}
J^{\mu}{}_{\sigma}= -\beta_1 \nabla^{\mu} u_{\sigma}-\beta_2
\, \delta^{\mu}_{\sigma} \, \nabla_{\rho} u^{\rho} -\beta_3 \nabla_{\sigma} u^{\mu}.
\nonumber
\end{equation}

Given Eqs.~\eqref{uvector} and \eqref{tmunu}, the nonvanishing components
of the background stress tensor are,
\begin{eqnarray}
T_{00}^{(u)} &=&  \beta_1 m^2\left({\dot{b}\over b}\right)^2,
\nonumber\\
T_{11}^{(u)}&=&T_{22}^{(u)}= \beta_1 m^2 a^2 \left({\dot{b}\over b}\right)^2,
\nonumber\\
T_{33}^{(u)} &=&  \beta_1 m^2 \left(  \dot{b}^2 -2 \ddot{b} b -4 {\dot{a}\dot{b} b \over a} \right).
\end{eqnarray}
The components of the energy-momentum tensor in our chosen
background are independent of $\beta_2$ and $\beta_3$.   For a non-trivial vector contribution to the stress-tensor (and consequently for an anisotropic metric resulting from the field equations), one must require that $\b_1 \neq 0$.  Note that the background solution does not satisfy the weak energy condition.

An inflating background solution requires,
\begin{equation}
\label{beta1bound}
8 \pi G m^2 \b_1 > - {1 \over 2}.
\end{equation}
This is not a  strong bound on $\b_1$ since we are  interested in backgrounds that have a small violation of isotropy, {\it i.e.}, $G m^2 << 1$.  

In this paper we study the classical stability of the homogeneous background solution to the equations of motion.  We expand the equations of motion to linear order about the homogeneous background. Since the background is homogeneous it is convenient to Fourier transform the small fluctuations  about the background in the co-moving spatial coordinates $(x_1,x_2,x_3)$ and examine the time dependence that the linearized equations of motion imply.  The  physical wave vectors ${\bar k}_i$ , $i=1,2$, and ${\bar k}_3$ are related to the co-moving wave vectors  by ${\bar k}_i=k_i/a(t)$ and ${\bar k}_3= k_3/b(t)$.

We begin with a flat-space stability analysis neglecting gravity since, when $\bar{k} \gg H$, such an analysis captures the essential physics.  Then we perform a stability analysis including gravity in two regions, $\bar k \gg H$ and $\bar k \ll H$.  We do not treat the transition region where the physical wavelengths of the modes cross the horizon.  In the final section of this paper  we  examine the flat-space energy of solutions to the equations of motion. Negative energy solutions indicate an instability that might not be evident from a study of the linearized equations of motion.


\section{Stability Analysis Neglecting Gravity}\label{nograv}
In the short-wavelength limit, $\bar k  \gg H$, it is physically intuitive that modes will not be able to resolve space-time curvature. Therefore we should be able to derive dispersion relations for small fluctuations in $u_\mu$ about a flat background space-time that will receive corrections suppressed by $G m^2$ when gravity is included.  In this case, we have the action,
\begin{eqnarray}
S &=& \int d^4x (  -\beta_1 (\pa^\mu u^\sigma) (\pa_\mu u_\sigma) - \beta_2 (\pa_\mu u^\mu)^2 \nonumber \\ &-& \beta_3 (\pa^\mu u^\sigma) (\pa_\sigma u_\mu) + \lambda(u^{\mu} u_{\mu} - m^2)).
\end{eqnarray}
The Euler-Lagrange equation for $\l$ gives, 
\begin{equation}
\label{flatfix}
\h_{\mu \nu}u^{\mu}u^{\nu}= m^2,
\end{equation}
while the equation of motion for the vector field gives,
\begin{equation}\label{eom}
\l u^\m + \b_1 \pa_\n \pa^\n u^\m + (\b_2+\b_3) \pa^\m (\pa_\n u^\n) = 0.
\end{equation}
We can use Eq.~(\ref{flatfix}) to solve for $\l$ by taking the inner product of the vector Euler-Lagrange equation \eqref{eom} with the vector,
\begin{equation}
\label{lambda}
\l = -\frac{\b_1}{m^2} u_\m \pa_\n \pa^\n u^\m - \left(\frac{\b_2+\b_3}{m^2} \right) u_\m \pa^\m (\pa_\n u^\n).
\end{equation}

We consider small perturbations about the background,  $ \bar{u}^\m = m \eta^{\m 3} $, $\bar{\l}=0$, and $\bar{g}_{\m\n} = \h_{\m\n}$.  We denote small perturbations about the background vector field by $v^\m$ so that, 
$$
u^\m = \bar{u}^\m + v^\m.
$$
Since we are neglecting gravity, all greek indices are raised and lowered by the Minkowski metric. The perturbation about the background value of the Lagrange multiplier, $ \d \l = \lambda - \bar{\l} $, is a function of the vector field perturbations through Eq.~\eqref{lambda},
\begin{equation}
m \d \l + (\b_2+\b_3) \pa_3 (\pa_\n v^\n) = 0. 
\end{equation}
From Eq.~\eqref{flatfix} we see that $v_3 = 0$. And by expanding  Eq.~\eqref{eom} to first order in perturbations we have the following equations for the non-trivial vector field perturbations ($i \in \lbrace 1, 2 \rbrace $):
\begin{eqnarray}
\label{EOM}
\b_1  \pa_\n \pa^\n v^0 + (\b_2+\b_3) \pa^0 (\pa_\n v^\n) &=& 0, \nonumber \\
\b_1  \pa_\n \pa^\n v^i + (\b_2+\b_3) \pa^i (\pa_\n v^\n) &=& 0.
\end{eqnarray}

We Fourier transform the components of $v^\m$, and in terms of the Fourier modes (transformed in both space and time),
\begin{eqnarray}
\label{newdecomp}
v^0(\o, \vec{k}) =  k_3 \theta_1 (\o, \vec{k}), \qquad \nonumber \\ 
\text{and} \qquad v^i (\o, \vec{k}) =  k_i \theta_2 (\o, \vec{k}) + \e_{i j} k_j \j (\o, \vec{k}),  
\end{eqnarray}

where $i, j \in \lbrace1,2 \rbrace$ and we have arranged for all scalar ($\theta$) and pseudoscalar\footnote{We use the name ``pseudoscalar'' for the $\psi$'s not because these objects are odd under parity but in order to distinguish them from the $\theta$'s. The two different kinds of scalars decouple.} ($\psi$) components components to have the same mass dimension.  With this decomposition, linear combinations of the Fourier transform of equations \eqref{EOM} become,
\begin{multline}
\label{sceqn1}
\left[\b_1 (\kp^2 + k_3^2) - (\b_1 + \b_2 +\b_3) \o^2 \right] k_3 \theta_1 \\ + (\b_2 + \b_3) \o \kp^2 \theta_2= 0, 
\end{multline}
\begin{multline}
\left[ \b_1 (\kp^2 + k_3^2 -\o^2) + (\b_2 +\b_3) \kp^2 \right] \theta_2 \\ -(\b_2 + \b_3) \o k_3 \theta_1 = 0, 
\end{multline}
\begin{equation}
\b_1 \kp^2 (\kp^2 + k_3^2 - \o^2) \j = 0,
\end{equation}
where $ \kp^2 \equiv \sum_{i=1}^2 k_i k_i$.

We first consider the case where $\b_1 + \b_2 +\b_3 \neq 0$. The scalar mode $\theta_1 = (\kp^2/ \o k_3) \theta_2$ propagates with the dispersion relation,
\begin{equation}
\o^2 = \kp^2 + k_3^2.
\end{equation}
The pseudoscalar mode also propagates with this dispersion relation.  The scalar mode $\theta_1 = (\o / k_3)\theta_2$ propagates with the dispersion relation,
\begin{equation}\label{shock dispersion}
\o^2 = \kp^2 + k_3^2\left(\frac{\b_1}{\b_1+\b_2 +\b_3}\right).
\end{equation}

When $\b_1 + \b_2 +\b_3 = 0$, Eq.~(\ref{sceqn1}) becomes a constraint equation and therefore the number of scalar modes is decreased by one while the pseudoscalar equation remains dynamical.  
A mode disappears at first order when $\b_1 + \b_2 +\b_3 = 0$ because there is an enhanced gauge symmetry. At first order in perturbations, the transformation $v_\m \rightarrow v_\m + \partial_\m \alpha(t, x_1, x_2)$ leaves the Lagrangian invariant.  The scalar function $\alpha$ is restricted to have no $x_3$ dependence at first order by the constraint $u_\m u^\m = m^2$.  The disappearing mode,
\begin{equation} \label{disappearing mode}
v^0(\o, \vec{k}) = \o \theta_2 (\o, \vec{k}), \qquad  v^i (\o, \vec{k})= k^i \theta_2(\o, \vec{k}), 
\end{equation}
satisfies (in coordinate space) $v_\m(t, \vec{x}) = \partial_\m \phi(t, \vec{x})$ for $\m = 0, 1, 2$. When $\b_2 + \b_3 = -\b_1$, Eqs.~\eqref{EOM} imply that $\phi(t, \vec{x})$ is independent of $x_3$. Thus the mode \eqref{disappearing mode} becomes a gauge artifact when $\b_1 + \b_2 + \b_3 = 0$.


\section{Stability Analysis Including Gravity}

In this analysis we have four  vector field fluctuations and ten metric fluctuations to consider. However, we can use  diffeomorphism invariance to remove some of them.  Under an infinitesimal coordinate transformation, $x^\mu \rightarrow x^\m + \xi^\m$, the vector field transforms to $u^{\lambda}+\Delta u^{\lambda}$ where,
\begin{equation}
\Delta u^{\lambda}=u^{\mu}\partial_{\mu} \xi^{\lambda} -\xi^{\mu}\partial_{\mu}u^{\lambda},
\end{equation}
while the metric transforms to $g_{\mu \nu}+\Delta g_{\mu \nu}$ where,
\begin{equation}
\Delta g_{\mu \nu}= -(\nabla_\m \xi_\n + \nabla_\n \xi_\m).
\end{equation}

We choose a gauge where  the first order fluctuations in the (contravariant) four-vector field about their background, $\bar{u}^\m =  \eta^{\m 3} m/b(t)$, vanish (\textit{i.e.}~$\d (u^\mu) = v^\m = 0$ where $u^\m = \bar{u}^\m + v^\m$). This gauge condition is satisfied by fixing, 
\begin{align*}
- v^\l &= \Delta u^\l \vert_{1^{st} \text{order}} 	\\
	&= \bar{u}^\m \partial_\m \xi^\lambda - \xi^\m \partial_\m \bar{u}^\l \\
	&= {m \over b(t)} \partial_3 \xi^\l + \delta^{\l}_{3} \xi^0 {m \over b(t)} H_b.
\end{align*} 

It is important to note that the gauge freedom has not been completely exploited. The condition $\d (u^\mu) = 0$ is invariant under gauge transformations of the form $x^\m \rightarrow x^\m + \zeta^\m$ where,
\begin{eqnarray}
\label{residual}
\zeta^\m = ( f_0(t,x_1,x_2), f_1(t,x_1,x_2), f_2(t,x_1,x_2), \nonumber \\
 f_3(t,x_1,x_2)- H_b x_3 f_0(t,x_1,x_2) ).
\end{eqnarray}
Once the above four functions of three coordinate variables are chosen, the gauge has been completely specified.  This residual gauge freedom will be used in the analysis of long-wavelength fluctuations.

The vector field satisfies the Lagrange multiplier equation, $u^2 = m^2$.  This equation and our choice of gauge imply that,
\begin{equation}
(\d g_{\m\n})u^\m u^\n + 2 g_{\m\n} (v^\m) u^\n  = (\d g_{33}) \frac{m^2}{b^2}   = 0,
\end{equation}
and thus we are left with nine metric fluctuations to consider.


Without loss of generality, we have chosen coordinates such that the background vector field lies entirely along the $x^3$ direction. Since isotropy is broken, only an SO(2) spatial symmetry remains. Let the Roman indices ${i, j, k}$ be SO(2) indices that run from $1$ to $2$. Define the `comoving' metric perturbations, $h_{\mu \nu}$, through the following equation,
\begin{eqnarray}
 \label{background}
ds^2 &=& - (1 + h_{00})\, dt^2 + 2 a(t) \,h_{0 i} \,dt dx^i \nonumber \\
&+& 2 b(t)\, h_{0 3} \, dt dx^3 +2 a(t) \,b(t) \,h_{i 3} dx^i dx^3  \\
&+& a(t)^2 \, (\delta_{i j} + h_{i j}) dx^i dx^j + b(t)^{2}(1+h_{33})(dx^{3})^{2} \nonumber
\end{eqnarray}
We work to first order in the small perturbations.  After making our gauge choice, there are nine perturbations, as well as nine independent equations for the system. One can show that the equations of motion for the vector field are equivalent to the energy-momentum conservation equations. Furthermore, energy-momentum conservation (and thus the equations of motion) follow from Einstein's equations. The identity to first order in perturbations is,
$$
8 \pi G \delta(\nabla_{\mu} T^{\mu}_{\nu}) =  \nabla^{\mu} \delta E_{\mu \nu},$$
where,
$$\delta E_{\mu \nu} =  \delta (G_{\mu \nu}) - 8 \pi G \delta (T_{\mu \nu} ). $$
Finally, only nine of Einstein's Equations are independent. The interdependence among Einstein's equations in our gauge can be found by working out the identity above for $\nu = 3$, since the left hand side vanishes identically,
$$
\left( \partial_t  + 2(H_a + H_b) \right) { \delta E_{03} \over b(t) }  =  \delta^{i j} { \partial_i \delta E_{j 3} \over a(t)^2 b(t)} + { \partial_3  \delta E_{3 3} \over b(t)^3}.
$$
Thus we take as our complete set of equations at first order in small perturbations all of Einstein's equations except for $\delta E_{33}$.

We Fourier transform,
$$
h_{\mu \nu}(t, k_1, k_2, k_3) = \int {d^3 x \over (2 \pi)^{3/2} } h_{\mu \nu}(t, x^1, x^2, x^3) e^{- i\vec{k} \cdot \vec{x}}, $$ and define,
$$
\bar{k}_i = k_i / a(t), \qquad \bar{k}_{\bot}^2 = \bar{k}_i \bar{k}_i, \qquad \text{and} \; \bar{k}_3 = k_3 / b(t), 
$$ where $i \in \{1,2\}$ and  summation is implied over repeated indices. Also define the constants,
\begin{equation}\label{lambdas}
\lambda_n = 8 \pi G m^2 \beta_n  \qquad \text{and} \qquad \Lambda = 8 \pi G \rho_{\Lambda}.
\end{equation}
Then the Fourier-transformed field equations become: 
\begin{widetext}
\begin{align}
\delta E_{00}  &=
\left( (\L + \bar{k}_3^2 (\l_1 + \l_2 + \l_3) \right) h_{00} + 2 i  \bar{k}_3 (\l_1 H_b - H_a  ) h_{03} - i (H_a + H_b) \bar{k}_i h_{0 i} - \bar{k}_3 \bar{k}_i h_{i 3} \nonumber \\
&+ \half \left( \bar{k}_{\bot}^2 + \bar{k}_3^2 (1 - 2 \l_2) +(H_a + H_b) \partial_t \right) h_{i i} - \half \bar{k}_i \bar{k}_j h_{i j} = 0,
\end{align}

\begin{align}
\frac{2 \delta E_{0 i}}{a(t)} &=
- i \bar{k}_i (H_a + H_b) h_{00} -\bar{k}_3 \bar{k}_i  h_{03}  + (2 \L - 2 H_a^2 - 2 H_a H_b - 2 H_b^2 (1 + \l_1) + \kpb^2 + \k3^2 (1 + 2 \l_1 + 2 \l_3) ) h_{0 i} \nonumber \\
 &-\ki \kj h_{0 j} + i \k3 ( -H_a + H_b(1 + 4 \l_1) + \partial_t )   h_{i 3} + i ( \kj \partial_t h_{i j} - \ki \partial_t h_{j j}  )= 0,
\end{align}

\begin{align}
\frac{\delta E_{03}}{b(t)}  &= i \bar{k}_3  ( \lambda_1 H_b - H_a ) h_{00} + \left(\L - 3 H_a^2 + \l_1(4 H_a H_b + H_b^2 + 2 \bar{k}_3^2) + \half (1 + 2 \l_1 - 2 \l_3) \kpb^2 \right) h_{03} \nonumber \\
&-\half \bar{k}_3 \bar{k}_i h_{0 i} + i \bar{k}_i \half(1 + 2 \l_1 - 2 \l_3)(H_a - H_b + \partial_t) h_{i 3} - i  \bar{k}_3  \half(H_a - H_b(1 + 2 \l_1) + \partial_t) h_{i i} =0,
\end{align}

\begin{align}
\frac{\delta E_{i j}}{a(t)^2} &=
\delta_{i j} \left(-(H_a^2 + H_a H_b + (1 + \l_1 ) H_b^2) + \half \kpb^2 + \half(1 - 2 \l_2) \k3^2 - \half (H_a + H_b) \partial_t \right) h_{00}  \nonumber \\
&- \half \ki \kj h_{00} +  i \k3  \d_{i j}(H_a + (1 + 2 \l_1) H_b + \partial_t) h_{03}  - i \bar{k}_{(i} (H_a + H_b + \partial_t ) h_{j) 0} + i \d_{i j} \bar{k}_l (H_a + H_b + \partial_t ) h_{0 l} \nonumber \\
  &+ \k3 ( \d_{i j} \bar{k}_{l}  h_{l 3}- \bar{k}_{(i}  h_{j) 3} ) - \half \d_{i j} (\k3^2(1 - 2 \l_2 ) + (2 H_a + H_b + \partial_t) \partial_t ) h_{k k}  \nonumber \\
  &+  \left( \L - (H_a^2 + H_a H_b + H_b^2 (1 + \l_1) ) + \half \k3^2 (1 + 2 \l_1 + 2 \l_3) + \half(2 H_a + H_b + \partial_t) \partial_t  \right) h_{i j} = 0,
\end{align}

\begin{align}
\frac{2 \delta E_{i 3}}{a(t) b(t)}  &= - \k3 \ki h_{00} - i \ki (1 + 2 \l_1 - 2 \l_3)(2 H_b + \partial_t ) h_{03}  - i \k3 (3 H_a - H_b(1 + 4 \l_1) + \partial_t) h_{0 i} \nonumber \\
&+  \left(2 \L + H_a^2(2 \l_1 - 2 \l_3 -5) + H_a H_b (1 + 10 \l_1 - 2 \l_3) - 2 H_b^2(1 + \l_1 - 2 \l_3)  
+ 4 \k3^2 \l_1\right) h_{i 3} \nonumber \\
&+ (1 + 2 \l_1 - 2 \l_3)\left( \delta_{ij}(\kpb^2 + (2 H_a + H_b + \partial_t)\partial_t) - \ki \kj \right) h_{j 3}+  \k3 (\ki h_{j j} - \kj h_{i j} ) = 0.
\end{align}
\end{widetext}


\subsection{Short-wavelength limit of the field equations}

Here we consider the limit  $(H / \bar{k} ) \rightarrow  0$, which corresponds to modes with physical wavelengths much shorter than the Hubble radius.  Such modes have periods much shorter than the Hubble time; therefore we can treat the $\bar{k}_i$ as constants  independent of time (\textit{i.e.}~$H_{a,b}$ are effectively zero in this limit).  The background solution in Eq.~\eqref{backgrounda} gives $\sqrt{\L} \sim H_b$ and thus we also take $(\L / \bar{k}^2 ) \rightarrow 0$. 

The various degrees of freedom in the field equations decouple into six equations governing the scalar modes and three equations governing the pseudoscalar modes if one uses the decomposition $(i,j=1,2)$:
\begin{eqnarray}
\label{decomp}
h_{00} &=& \kpb^2 \bar{\Theta}_1,\nonumber \\ 
h_{0i} &=& \bar{k}_3 \left(\bar{k}_i \bar{\Theta}_2 + \e_{ij} \bar{k}_j \bar{\Psi}_1 \right), \nonumber \\ 
h_{03} &=& \bar{k}_3^2 \bar{\Theta}_3,   \\
h_{ij} &=& \d_{ij} \frac{\kpb^2}{2} \bar{\Theta}_4 + \left[\bar{k}_i\bar{k}_j -\d_{ij} \frac{\kpb^2}{2}\right] \bar{\Theta}_5 \nonumber \\ &+& \bar{k}_{(i} \e_{j)k} \bar{k}_k \bar{\Psi}_2, \nonumber \\
h_{i3} &=& \bar{k}_3 \left(\bar{k}_i \bar{\Theta}_6 + \e_{ij} \bar{k}_j \bar{\Psi}_3\right).  \nonumber
\end{eqnarray} 
Note that all of the scalar ($\bar{\Theta}$) and pseudoscalar ($\bar{\Psi}$) components have the same mass dimension.  We make the ansatz for the time dependence of the fields above,  
$$
\bar{\Theta}_l(t, \vec{k}) = \bar{\Theta}_l (\o, \vec{k}) e^{i \o t}~~ \text{and}~~  \bar{\Psi}_l(t, \vec{k}) = \bar{\Psi}_l (\o, \vec{k}) e^{i \o t}.
$$

The decomposition \eqref{decomp} aids in solving the field equations in the usual way: linear combinations of Einstein's equations that resemble the above decomposition lead to equations coupling only (pseudo)scalar fields.

There are six scalar fields in the mode decomposition (\ref{decomp}).  The functional and algebraic constraints on these fields are given by the field equations:
\begin{widetext}
\begin{eqnarray*}
0 &=& \frac{\delta E_ {00}}{\kpb^2} = \bar{k}_3^2 (\l_1 + \l_2+\l_3)\bQ_1 +  \left[\frac{\kpb^2}{4}+\frac{\bar{k}_3^2}{2}(1-2\l_2)\right] \bQ_4 - \frac{\kpb^2}{4} \bQ_5 - \bar{k}_3^2 \bQ_6,  \\
0 &=& \frac{2  \bar{k}_ i \delta E_ {0i}}{a(t) \bar{k}_3 \kpb^2}  = \bar{k}_3^2 (1 + 2\l_1+2\l_3)\bQ_2  - \bar{k}_3^2 \bQ_3  +\frac{\o}{\bar{k}_3}\left( \frac{\kpb^2}{2} \bQ_4  - \frac{\kpb^2}{2}\bQ_5  -\bar{k}_3^2 \bQ_6 \right), \\
0 &=& \frac{\delta E_ {03}}{b(t) \bar{k}_3^2}  = -\frac{\kpb^2}{2} \bQ_2 + \left[\frac{\kpb^2}{2}(1+2\l_1 - 2\l_3) + 2\bar{k}_3^2 \l_1\right] \bQ_3 +  \frac{\o}{\bar{k}_3} \left(\frac{\kpb^2}{2}\bQ_4 - \frac{\kpb^2}{2}(1+2\l_1 - 2\l_3) \bQ_6 \right), \\
0 &=& \frac{\delta E_ {ii}}{a(t)^2 \kpb^2}  = \left[ \frac{\kpb^2}{2}+\bar{k}_3^2(1-2\l_2)\right] \bQ_1 - \bar{k}_3 \o \bQ_2 - 2\frac{\o \bar{k}_3^3}{\kpb^2} \bQ_3 + \left[\frac{\o^2}{2} +\bar{k}_3^2 (\l_1 +2\l_2 +\l_3 - {1 \over 2})\right] \bQ_4 + \bar{k}_3^2 \bQ_6, \\
0 &=&\frac{ \bar{k}_i \bar{k}_ j \delta E_ {ij}}{a(t)^2 \kpb^4}  = \frac{\bar{k}_3^2}{2}(1-2\l_2) \bQ_1 - \frac{\o \bar{k}_3^3}{\kpb^2} \bQ_3 + \left[ \frac{\o^2}{4} - \frac{\bar{k}_3^2}{4} (1-2\l_1 -4\l_2 - 2\l_3)\right] \bQ_4 + \left[\frac{\bar{k}_3^2}{4} (1+2\l_1 + 2\l_3) - \frac{\o^2}{4}\right] \bQ_5,   \\
0 &=& \frac{ \bar{k}_i \delta E_ {i3}}{a(t) b(t) \bar{k}_3 \kpb^2}  = \frac{\o \bar{k}_3}{2} \bQ_2-\frac{\kpb^2}{2} \bQ_1  + \frac{\o \bar{k}_3}{2}(1+2\l_1 -2\l_3)\bQ_3 + \frac{\kpb^2}{4}( \bQ_4 -  \bQ_5) - \left[\frac{\o^2}{2}\left(1+2\l_1 - 2\l_3\right) - 2 \l_1 \bar{k}_3^2 \right] \bQ_6.
\end{eqnarray*}
\end{widetext}

Let us consider the dynamical degrees of freedom. We shall schematically denote $\o \bQ(\o, \vec{k}) \sim \partial_t \bQ(t, \vec{k}) \sim \dot{\bQ}(t) $. The first equation is a constraint. When $\b_1 + \b_2 + \b_3 \neq 0$,\footnote{Recall the definition, $\l_n \equiv 8 \pi G m^2 \b_n$. }  it can be solved to give $\bQ_1(t) = \bQ_1(\bQ_4(t), \bQ_5(t),\bQ_6(t))$. The next two equations can be solved to give $\bQ_2(t) = \bQ_2(\dot{\bQ}_4(t), \dot{\bQ}_5(t),\dot{\bQ}_6(t))$ and  $\bQ_3(t) = \bQ_3(\dot{\bQ}_4(t), \dot{\bQ}_5(t),\dot{\bQ}_6(t))$. In the last three equations, $\bQ_2$ and $\bQ_3$ appear only as $\dot{\bQ}_2(t)$ and $\dot{\bQ}_3(t)$. Thus, after making the substitutions for $\bQ_1$, $\bQ_2$, and $\bQ_3$, the last three equations form a system of second order ordinary differential equations of three functions. Thus, when $\b_1 + \b_2 + \b_3 \neq 0$, we expect three distinct dynamical modes.

When $\b_1 + \b_2 + \b_3 = 0$, the function $\bQ_1$ drops out of the constraint equation. Then the first equation can be solved for $\bQ_4(\bQ_5(t), \bQ_6(t))$ and the next two for  $\bQ_2(t) = \bQ_2(\dot{\bQ}_5(t),\dot{\bQ}_6(t))$ and  $\bQ_3(t) = \bQ_3(\dot{\bQ}_5(t),\dot{\bQ}_6(t))$. Since $\bQ_1$ enters the above equations with no time derivatives, one of the last three equations can be used to eliminate $\bQ_1$, and the remaining two become second order differential equations in time of two functions, $\bQ_5(t)$ and $\bQ_6(t)$.  Thus, when $\b_1 + \b_2 + \b_3 = 0$, we expect only two distinct dynamical modes in the scalar sector.

Here we present the propagating modes and their dispersion relations. The first scalar mode relates the amplitudes of the various fields by,
\begin{equation}
\bQ_1 = {\bar{k}_3 \over \o} \bQ_2 = \bQ_4 = {\kpb^2 \bQ_5 \over 2\o^2 - \kpb^2 } \text{;} \qquad \bQ_3 = 0 = \bQ_6,
\end{equation}
and propagates with dispersion relation,
\begin{equation}
\label{sc1}
\o^2 = \kpb^2 + \bar{k}_3^2(1+2\l_1 +2 \l_3).
\end{equation}
The next scalar mode is characterized by the amplitude relationships,
\begin{eqnarray}
\bQ_1 = \bQ_4 = \bQ_5 &=& {2 \bar{k}_3 \o \over \o^2 + \kpb^2} \bQ_2 \nonumber \\ 
= {2 \bar{k}_3 \o \over \kpb^2 (1+ 2\l_1 +2\l_3)}\bQ_3 &=& {2  \over 1+2\l_1 +2\l_3} \bQ_6,
\end{eqnarray}
and propagates with dispersion relation,
\begin{equation}
\label{sc2}
\o^2 = \kpb^2 + \bar{k}_3^2\left(\frac{\l_1(1+2\l_1 +2 \l_3)}{\l_1+\l_1^2-\l_3^2}\right).
\end{equation}
The last propagating scalar mode only occurs when $\b_1 + \b_2 + \b_3 \neq 0$ and is characterized by the amplitude relationship,
\begin{eqnarray}
\bQ_5 &=& {\bar{k}_3 \over \o} \bQ_2  = {2 \bar{k}_3 (1 + \l_1)  \over \o(1+2\l_1 +2 \l_3)} \bQ_3 \nonumber \\
&=& {2 (1 + \l_1)  \over 1+2\l_1 +2 \l_3} \bQ_6 = \frac{\kpb^2 }{\left( \o^2+ {\bar{k}_3^2 \l_1 (1+2\l_1 +2 \l_3) \over 1 + \l_1} \right)}\bQ_1 \nonumber \\
&=& \frac{\kpb^2 }{\left(\kpb^2 -2 {\bar{k}_3^2 \l_1(1+2\l_1 +2 \l_3) \over 1 + \l_1} \right)} \bQ_4, 
\end{eqnarray}
and dispersion relation,
\begin{multline}
\label{sc3}
\o^2 = \kpb^2 \\+ \bar{k}_3^2\left(\frac{\l_1(1+2\l_1 +2 \l_3)}{(1+\l_1)}\right)\left(\frac{1-\l_1 - 3\l_2 - \l_3}{\l_1+\l_2 +\l_3}\right). 
\end{multline}
This last mode corresponds to the flat-space mode \eqref{disappearing mode}.
As we found in the flat-space analysis, this mode is absent when $\b_1 + \b_2 + \b_3 = 0$. (Note that in the short-wavelength limit, $a(t)$ and $b(t)$ are treated as constants, so covariant derivatives become ordinary derivatives and thus $\b_1 + \b_2 + \b_3 = 0$ corresponds to the case where the Lagrange density takes the form of a field strength squared---as in the flat-space analysis.)    

For general Fourier components, we can guarantee that $\o^2 \geqslant 0$ if and only if the coefficient of the $\bar{k}_3^2$ term is positive semidefinite.  Thus we see that the stability of the modes gives the following conditions on the $\l_n$'s (and thus, recalling Eq.~\eqref{lambdas}, the $\b_n$'s):
\begin{eqnarray}
\l_1 + \l_3 \geqslant - {1 \over 2}, \nonumber \\
\frac{\l_1(1+2\l_1 +2 \l_3)}{\l_1+\l_1^2-\l_3^2}  \geqslant 0,\\
\frac{\l_1(1+2\l_1 +2 \l_3)}{(1+\l_1)}\left(\frac{1-2\l_2}{\l_1+\l_2 +\l_3} -1\right) \geqslant 0. \nonumber
\end{eqnarray}
We see that only the last equation gives us a condition on $\b_2$.  In the limit that $G m^2 << 1$, we have the following condition,
\begin{equation}
\frac{\l_1}{\l_1+\l_2 +\l_3} =  \frac{\b_1}{\b_1+\b_2 +\b_3} \geqslant 0,
\end{equation}
which is identical to what we found in the flat-space analysis. The other two constraints are trivially satisfied in this limit.  

The following equations for the pseudoscalar fields follow from the field equations:
\begin{eqnarray*}
0 &=& \frac{2 \e_{ij} \d E_{0i} \bar{k}_j}{\bar{k}_3 \kpb^2 a(t)} =  \left[\kpb^2+\bar{k}_3^2(1+2\l_1 +2\l_3)\right] \bJ_1 \\
 &-& \frac{\o}{\bar{k}_3} \left(\frac{\kpb^2}{2} \bJ_2 + \bar{k}_3^2 \bJ_3 \right),  \\ 
0 &=& \frac{\e_{lj} \bar{k}_l \bar{k}_i\d E_{ij}}{\kpb^4 a(t)^2} = -\frac{\o \bar{k}_3}{2} \bJ_1 +\frac{\bar{k}_3^2}{2} \bJ_3 \\  
&+& \left[\frac{\o^2}{4}-\frac{\bar{k}_3^2}{4}(1+2\l_1 + 2\l_3) \right] \bJ_2,  \\
0 &=& \frac{ \e_{ij} \d E_{i3} \bar{k}_j}{\bar{k}_3 \kpb^2 a(t) b(t)} = \frac{\o \bar{k}_3}{2} \bJ_1 -\frac{\kpb^2}{4} \bJ_2 \\
&-& \left[\frac{\o^2 - \kpb^2}{2}\left(1+2\l_1 - 2\l_3\right) - 2 \l_1 \bar{k}_3^2 \right] \bJ_3,
 \nonumber
\end{eqnarray*}
where  $\l_n = 8 \p G m^2 \b_n$. Here, the first equation is a constraint and there are two distinct dynamical modes.
The pseudoscalar eigenmode $  (2 \o \bar{k}_3 / \kpb^2) \bJ_1 = \bJ_2$, $\bJ_3=0$  propagates with the dispersion relation,
\begin{equation}
\label{ps1}
\o^2 = \kpb^2 + \bar{k}_3^2(1+2\l_1 +2 \l_3).
\end{equation}
The second pseudoscalar eigenmode is given by $(2 \bar{k}_3/\o)(1+ 2\l_1 +2 \l_3) \bJ_1 =  (1+ 2\l_1 +2 \l_3) \bJ_2 = 2 \bJ_3$ and propagates with the dispersion relation,
\begin{equation}
\label{ps2}
\o^2 = \kpb^2 + \bar{k}_3^2\left(\frac{\l_1(1+2\l_1 +2 \l_3)}{\l_1+\l_1^2-\l_3^2}\right).
\end{equation}
Note that the value of $\b_2$ does not affect the stability of any of the propagating pseudoscalar modes.  Also, the two pseudoscalar dispersion relations were already found to characterize two of the propagating modes in the scalar sector. 

When $\b_1 = - \b_3$ and $\b_2 = 0$ (the Maxwell case), all modes propagate with the speed of light. However, all modes would not necessarily propagate with the speed of light if we only required $\b_1 + \b_2 + \b_3 = 0$.

In the limit $Gm^2<< 1$ (or equivalently $m^2 << M_p^2$, where $M_p$ is the Planck mass) we have $\l_{n} <<1$.  The flat-space analysis considered earlier is insensitive to $\co(\l_{n})$ corrections to the dispersion relations and (of course) neglects gravitational degrees of freedom.  Here we show that the more general analysis simplifies to the flat-space analysis.
Since, in our gauge, $\d (u_{\mu}) = \d g_{\mu \nu} \bar{u}^{\nu} =  \left(m / b \right) \d g_{\mu 3}$, comparing Eq.~\eqref{newdecomp} and Eq.~\eqref{decomp} we expect that,
\begin{equation}\label{expected}
\theta_1 \propto \bQ_{3}, \qquad \theta_2 \propto \bQ_6, \qquad \text{and} \; \psi \propto \bJ_3.
\end{equation}
In the $\l_n << 1$ limit, the dispersion relation for the scalar mode, (\ref{sc2}), and the pseudoscalar mode, (\ref{ps2}), has the first order approximation,
\begin{equation}
\o^2 = \kpb^2 + \bar{k}_3^2 \left[1+\left(\frac{(\l_1 +\l_3)^2}{\l_1} \right) \right], \qquad 
\end{equation}
with amplitude relationship $\left(\bar{k}_3 \o / \kpb^2 \right)\bQ_3 = \bQ_6$.  
The dispersion relation in Eq.~(\ref{sc3}) has the first order approximation,
\begin{equation}
\o^2 = \kpb^2 + \bar{k}_3^2\left(\frac{\b_1}{\b_1+\b_2 +\b_3}\right) \left[ 1 - 3\l_2 + \l_3 \right] ,
\end{equation}
with amplitude relationship $\left(\bar{k}_3 / \o \right) \bQ_3 =  \bQ_6$.  

The $\l_n \rightarrow 0$ limit of dispersion relations \eqref{sc2}, \eqref{sc3}, and \eqref{ps2} and the corresponding mode amplitude relationships, along with Eq.~\eqref{expected}, lead to the propagating modes and dispersion relations found in section \ref{nograv}. (In the flat-space analysis, one can rescale the Fourier modes in order to put them in the form of $\kpb$ and $\bar{k}_3$ since this simply results in an overall constant rescaling of the integration measure.)
We found only three (two when $\b_1 + \b_2 + \b_3 = 0$) distinct modes in the analysis neglecting gravity because the $\bQ_6 = 0 = \bQ_3$ and $\bJ_3 = 0$ modes are purely gravitational.  The purely gravitational modes 
have the following exact dispersion relation,
\begin{equation}
\o^2 = \kp^2 + k_3^2 \left( 1+2 \l_1 + 2\l_3 \right).
\end{equation}
As $\l_n \rightarrow 0$, the gravitational modes have the usual graviton dispersion relation (as expected).  
We see that the intuitive flat-space analysis is reproduced by the more general analysis involving coupling to gravity.

In  \cite{Lim:2004js}, Lim carried out a similar perturbative analysis of a fixed-norm, timelike, Lorentz-violating vector field in a de-Sitter background.  Lim considers the model of Ref.~\cite{Carroll:2004ai} which is a slight simplification of the model in Ref.~\cite{Jacobson:2000xp}.\footnote{The action in \cite{Jacobson:2000xp} includes a term quartic in the vector field.}  He found rescaled mode propagation speeds very similar to those that we found when $\b_1 + \b_2 + \b_3 \neq 0$. There is a one-to-one correspondence of the inverse of his propagation speeds with our $x_3$-direction propagation speeds when $m^2 \rightarrow - m^2$.


\subsection{Long-wavelength limit of the field equations}
We now consider the behavior of the modes  after they cross the Hubble Horizon and work in the limit where ${\bar  k} / H \ll 1$. First simplifying Einstein's equations in this limit and then performing a mode decomposition yields simple differential equations for the various modes. 
The  field equations decouple into six equations governing the scalar ($\Theta$) modes and three equations governing the pseudoscalar ($\Psi$) modes if one uses the decomposition $(i,j=1,2)$:
\begin{eqnarray}
\label{decomp2}
h_{00} &=&  \Theta_1,\nonumber \\ 
h_{0i} &=& {1 \over \kp} \left(k_i \Theta_2 + \e_{ij} k_j \Psi_1 \right), \nonumber \\ 
h_{03} &=& \Theta_3,  \\
h_{ij} &=& \frac{\d_{ij}}{2}\Theta_4 + \left[{k_ik_j \over \kp^2} -\frac{\d_{ij}}{2}\right] \Theta_5 + \frac{k_{(i} \e_{j)k} k_k}{\kp^2} \Psi_2,  \nonumber \\
h_{i3} &=& {1 \over \kp} \left(k_i \Theta_6 + \e_{ij} k_j \Psi_3\right),  \nonumber 
\end{eqnarray} 
where $\kp = \sqrt{k_1^2 + k_2^2}$.  Note that all of the scalar ($\Theta$) and pseudoscalar ($\Psi$) components have the same mass dimension and are zeroth order in $\kp, k_3$.  

The scalar mode Einstein equations are:
\begin{eqnarray*}
0 &=& H_b (3 + 4 \l_1) \Theta_1(t) + \partial_t \Theta_4(t),\\
0 &=& 4 H_b (1 + \l_1)(H_b (3 + 4 \l_1) + \partial_t) \Theta_1(t) \\
&+& \left( (3 + 4 \l_1) H_b \partial_t + \partial_t^2 \right) \Theta_4(t) ,\\
0 &=& \left(H_b(3 + 4 \l_1) \partial_t +  \partial_t^2 \right) \Theta_5(t),\\
0 &=& (2 H_b^2 \l_1 (3 + 2 \l_1) + H_b (3 + 4 \l_1) \partial_t + \partial_t^2 ) \Theta_6(t),
\end{eqnarray*}
where we have used the background solution, $H_a = H_b (1 + 2 \l_1)$ and $\L = H_b^2 (1 + \l_1)(3 + 4 \l_1)$. 
The other two scalar equations are trivially zero. This means that $\Theta_2(t)$ and $\Theta_3(t)$ are unconstrained functions of time. The solution for the remaining scalar modes is:
\begin{align}
\Theta_1(t) & = \Theta_1(0) e^{-t H_b (3 + 4 \l_1)}, \label{Q1soln}\\
\Theta_4(t) & =  \Theta_1(0) \left(e^{-t H_b (3 + 4 \l_1)} - 1 \right) + \Theta_4(0),  \label{Q4soln} \\
\Theta_5(t) & = \Theta_5(0) e^{-t H_b (3 + 4 \l_1)} + A_5(e^{-t H_b (3 + 4 \l_1)} - 1),  \label{Q5soln} \\
\Theta_6(t) & =  e^{-2 t H_b \l_1}\left( \Theta_6(0) e^{-3 t H_b} + A_6 (e^{-3 t H_b} - 1)\right),  \label{Q6soln}
\end{align}
where $A_4$ and $A_5$ are arbitrary constants. One should note that $H_a - H_b = 2 \l_1 H_b$ and $H_b (3 + 4 \l_1) = 2 H_a + H_b$.

The nontrivial psuedoscalar mode Einstein equations imply,
\begin{eqnarray*}
0 &=& \left( H_b (3 + 4 \l_1) \partial_t + \partial_t ^2 \right) \Psi_2(t),\\
0 &=& \left( 2 H_b^2 \l_1 (3 + 2 \l_1) + H_b (3 + 4 \l_1) \partial_t + \partial_t^2 \right) \Psi_3(t),
\end{eqnarray*}
where we have again used the background solution, Eq.~\eqref{background}. The third equation vanishes identically, so there are no constraints on $\Psi_1(t)$. 

The solutions for the other modes are,
\begin{align}
\Psi_2(t) &= \Psi_2(0) e^{-t H_b (3 + 4 \l_1)} + B_2 \left( e^{-t H_b (3 + 4 \l_1)} - 1 \right),  \label{J2soln}\\
\Psi_3(t) &= e^{-2 t H_b \l_1} \left( \Psi_3(0) e^{-3 t H_b} + B_3 (e^{-3 t H_b} - 1) \right),   \label{J3soln}
\end{align}
where $B_2$ and $B_3$ are arbitrary constants. 

The unconstrained scalar and pseudoscalar modes, $\Q_2(t)$, $\Q_3(t)$, and $\Psi_1(t)$, are, unsurprisingly, gauge artifacts. Importantly, $\J_3(t)$ is unchanged under any transformation of the form \eqref{residual}. We show this explicitly in the appendix.

That  $\J_3(t)$ is not a gauge artifact implies that $\b_1$ (recalling that $\l_1 = 8 \pi G m^2 \b_1$) must be greater than zero, 
\begin{equation}
\label{beta1 constraint} 
\b_1 > 0, 
\end{equation} 
in order for this mode to decay.   Thus, assuming Eq.~\eqref{beta1 constraint}, no comoving modes grow in the limit $k / H << 1$.   

For the timelike vector \cite{Lim:2004js}, the $\b_1 > 0$ bound guarantees a positive definite  flat-space Hamiltonian at second order in perturbations; however, $\b_1 > 0$ is not a requirement for lowest order perturbative stability.  For the case of a timelike vector in de-Sitter space, all comoving modes decay as $ 1 / a(t) $  \cite{Lim:2004js}; however, for a spacelike vector we find that $\Q_6(t)$ and $\J_3(t)$ fall off more slowly than $ 1 / a(t) $ while $\Q_1(t)$, $\Q_4(t)$, $\Q_5(t)$ and $\J_2(t)$ decay more quickly. (Here we assume that $H_a - H_b = 2 H_b \l_1$ is smaller than $H_a $.)


\section{Flat-Space Energies}

In the previous sections we have studied the classical equations of motion linearized about the homogeneous background solution. For small $Gm^2$ they were stable provided $\beta_1 >0$ and $\beta_1+\beta_2+\beta_3 \ge 0$.  In this section we examine the classical energy of solutions to the flat-space equations of motion. Negative energy solutions indicate an instability that might not have shown up in our study of the linearized equations of motion.

In flat-space one can write the vector field action as,
\begin{multline}\label{action}
S_u =  - \int d^4 x \Bigl( {\b_1 \over 2} F_{\m\n}F^{\m\n} + \left(\b_1 + \b_2 + \b_3 \right) \left( \partial_\m u^\m \right)^2 \\
- \l (u_\m u^\m - m^2) \Bigr),
\end{multline}
where $F_{\m \n} = \partial_\m u_\n - \partial_\n u_\m$  and boundary terms are set to zero.  Let $i,j \in \{ 1,2 \}$ as in previous sections. The energy functional derived from the above action is,\footnote{The Lagrange multiplier constraint is holonomic. Whether a holonomic constraint is imposed before or after forming the stress-energy tensor has no effect on the tensor's final form.} 
\begin{multline}
\label{FullH}
E_u = \int d^3 x \, { T^{(u) 0} }_0 =  \int d^3x \left( { \delta \lag_u \over \delta (\partial_0 u_\rho) } \partial_0 u_\rho - \lag_u \right) \\
= \b_1 \int d^3x \left({1 \over 2} F_{\m \n}F_{\m \n} + 2 u_0  \partial_\m F_{\m 0} \right) \\
-\left( \b_1 +\b_2 + \b_3 \right)  \int d^3x \left( \left( \partial_0 u_0 \right)^2 - \left( \partial_i u_i + \partial_3 u_3 \right)^2 \right) \\
-  \int d^3x \l (u_\m u^\m - m^2) ,  
\end{multline}
where, in the last expression, all indices are intentionally lower and repeated indices should be summed without factors of the metric.

We take $u_\m = ( v_0, v_1, v_2, m + v_3 ) $ as in section \ref{nograv}. Physical field configurations must satisfy the Lagrange multiplier equation of motion, $u^2 = m^2$, which implies,
\begin{equation}
v_3 = m \left( \sqrt{1+ \frac{v_0^2 - v_i^2}{m^2}} -1 \right) =  \frac{v_0^2 - v_i^2}{2m} + \ldots.
\end{equation}
Thus $v_3$ is actually second order in perturbations. This expansion also makes manifest that an expansion order-by-order in perturbations is equivalent to an expansion in powers of $m^{-1}$. 

Imposing only the Lagrange multiplier equation of motion, the lowest order ($\co \left( m^0 \right)$) piece of Eq.~\eqref{FullH} is,
\begin{multline}
\label{Hquad}  
E_u^{(0)} = \b_1 \int d^3x \Bigl(  {1 \over 2} F_{ij}F_{ij} - (\partial_3 v_0)^2 +(\partial_3 v_i)^2  \\ 
- ( \partial_i v_0 +  \partial_0 v_i ) F_{i 0} \Bigr)  \\
-\left( \b_1 +\b_2 + \b_3 \right)  \int d^3 x \left( \left( \partial_0 v_0 \right)^2 - \left( \partial_i v_i \right)^2 \right),
\end{multline} since $v_3 = \co (m^{-1})$.
When $\b_1 + \b_2 +\b_3 \neq 0$, we find that the mode described by the dispersion relation (\ref{shock dispersion}) has energy  density (at lowest order in perturbations), 
$$
{E_u^{(0)} / V} =- 2 \b_1 \o^2 k_3^2 \tilde{\theta}_2^2,
$$ while the modes that propagate with dispersion relation $ \o^2 = \kp^2 + k_3^2 $ have energy density,
$$
{E_u^{(0)} / V} =2 \b_1 \o^2  \kp^2 \tilde{\psi}^2 \qquad \text{and} \qquad {E_u^{(0)} / V} = 2 \b_1 k_3^2 \kp^2 \tilde{\theta}_2 ^2 ,
$$ where $\tilde{\psi}$ and $\tilde{\theta}_2 $ are coordinate-independent coefficients of the plane wave solutions and $V$ is the volume of space. Thus for $\b_1 > 0$ the mode with dispersion relationship \eqref{shock dispersion} has negative energy while the others have positive energy.  The existence of a negative energy mode implies that the background field configuration, $\bar{u}^\m = m \eta^{\m 3}$, is unstable when  $\b_1 + \b_2 +\b_3 \neq 0$. 

However, as discussed in section \ref{nograv}, when $\b_1 + \b_2 +\b_3 = 0$, the mode that we have just shown to have negative energy (when $\b_1 > 0$) vanishes. Indeed, if $\b_1 + \b_2 +\b_3 = 0$, then the lowest order energy functional,
\begin{multline}
{E}^{(0)}_u =  \b_1 \int d^3x ( F_{0i} F_{0i}+ {1 \over 2} F_{ij}F_{ij} \\ + (\partial_3 v_0)^2 +(\partial_3 v_i)^2 ),
\end{multline}
is manifestly positive.


\section{Concluding Remarks}

We have studied the small fluctuations about a spatially homogeneous anisotropic inflationary background. The anisotropy was caused by a dynamical four-vector that was constrained to have a spacelike invariant norm. 

From the first order stability analysis of the equations of motion we derived the constraints $\b_1 > 0$ and $\b_1 + \b_2 + \b_3 \ge 0$ for the parameters in the vector Lagrangian \eqref{ulag}.  Moreover, we find that all modes have positive energy in flat-space and do not grow with time in the case where the kinetic term for the four-vector corresponds to a field strength tensor squared (\textit{i.e.}~for $\b_1 + \b_2 +\b_3 =0$).  A negative energy mode propagates if  $\b_1 + \b_2 +\b_3 \neq 0$, which implies that the background given by Eqs.~\eqref{uvector}-\eqref{backgrounda} is unstable when $\b_1 + \b_2 + \b_3 \ne 0$.

\subsection*{Post-submission note}

After the posting of this paper, in Ref.~\cite{Himmetoglu:2008zp}, an instability in the theory with a fixed-norm spacelike vector field with a field strength squared kinetic term was found  when the physical momentum of perturbations is comparable to the Hubble scale. We considered only physical momenta that were much greater than or much less than the Hubble scale.

\appendix*
\section{Gauge Artifacts}
A gauge transformation of the form in Eq.~\eqref{residual} induces the transformation $h_{\mu \nu}(t, \vec{k}) \rightarrow h_{\mu \nu}(t, \vec{k}) + \D h_{\mu \nu}(t, \vec{k})$ in the comoving Fourier-transformed metric perturbations. Explicitly, (i, j = 1, 2)
\begin{align}
\D h_{0 0}(t, \vec{k}) &= 2 \partial_t f_0(t, \kpv), \nonumber \\
\D h_{0 i}(t, \vec{k}) &= i {k_i \over a(t)} f_0(t, \kpv) - a(t) \pa_t f_i(t, \kpv), \nonumber \\
\D h_{0 3}(t, \vec{k}) &= b(t) \left( i \d{'}(k_3) \pa_t f_0(t, \kpv)- \pa_t f_3(t, \kpv) \right), \nonumber \\
\D h_{i j}(t, \vec{k}) &= - 2 \left( H_a \d_{i j} f_0(t, \kpv) +  i k_{(i} f_{j)}(t, \kpv)\right), \nonumber \\
\D h_{i 3}(t, \vec{k}) &= - {b(t) \over a(t)}  k_i \left( \d{'}(k_3)f_0(t, \kpv) + i f_3(t, \kpv ) \right), \nonumber
\end{align} 
where $\kpv = (k_1, k_2)$ and $f_\mu(t, \kpv)$ are the Fourier transforms of $f_\mu(t, x_1, x_2)$.  The corresponding transformations of the fields as defined in Eq.~\eqref{decomp2}  are, 
\begin{align}
\D \Q_1 &= \D h_{00} =  2 \partial_t f_0(t, \kpv) \label{deltaQ1},\\
\D \Q_2 &= {k_i \over \kp} \D h_{0 i} =  i {\kp \over a(t)} f_0(t,\kpv)- a(t) {k_i \over \kp} \pa_t f_i(t, \kpv) \label{deltaQ2},\\
\D \Q_3 &= \D h_{03} =b(t) \left( i \d{'}(k_3) \pa_t f_0(t, \kpv)- \pa_t f_3(t, \kpv) \right) \label{deltaQ3}
\end{align}
\begin{align}
\D \Q_4 &= \D h_{i i} = - 2 \left( H_a 2 f_0(t, \kpv) +  i k_{i} f_{i}(t, \kpv)\right) \label{deltaQ4}, \\
\D \Q_5 &= \left( {k_i k_j \over \kp^2} - {1 \over 2} \d_{i j}\right)\D h_{i j} = - i k_i f_{i}(t, \kpv) \label{deltaQ5}, \\
\D \Q_6 &= {k_i \over \kp} \D h_{i 3} \nonumber \\
	&= - {b(t) \over a(t)}  \kp \left( \d{'}(k_3)f_0(t, \kpv) + i f_3(t, \kpv ) \right)\label{deltaQ6},
\end{align}
and,
\begin{align}
\D \J_1&= {\e_{i j} k_j \over \kp} \D h_{0 i} =  - a(t) {\e_{i j} k_j \over \kp} \pa_t  f_i(t, \kpv), \label{deltaJ1}\\
\D \J_2 &= {k_{(i} \e_{j) k} k_k \over \kp^2} \D h_{i j} = i \e_{i j} k_i f_j(t, \kpv),  \label{deltaJ2}\\
\D \J_3 &= {\e_{i j} k_j \over \kp} \D h_{i 3} = 0.  \label{deltaJ3}
\end{align}

We see that $\J_3(t)$ is invariant under the residual gauge transformations in Eq.~\eqref{residual}. From Eqs.~\eqref{deltaQ2}, \eqref{deltaQ3}  and \eqref{deltaJ1}, it is clear that one can gauge away $\Q_3(t)$ by fixing $\pa_t f_3(t, \kpv)$ and $\pa_t f_0(t, \kpv) = 0$, and one can gauge away $\Q_2(t)$ and $\J_1(t)$ by fixing $\pa_t k_i f_i(t, \kpv)$, and $\pa_t \e_{i j} k_j f_i(t, \kpv)$, respectively. 




\begin{acknowledgments}

We thank Sean Carroll, Ted Jacobson, and Lotty Ackerman for their comments during the preparation of this manuscript.  This work was supported in part by DOE grant number DE-FG03-92ER40701. 
\end{acknowledgments}


\bibliography{references}

\end{document}